\documentclass[a4paper,aps,prl,twocolumn,showpacs,superscriptaddress,longbibliography,titlepage]{revtex4-1}
\usepackage{graphicx}
\usepackage{color}
\usepackage{bm}
\usepackage{bbm}
\usepackage{amsmath}
\usepackage{amsfonts}
\usepackage{lipsum}
\usepackage{hyperref}
\usepackage{dsfont}

\newcommand{\reffig}[1]{Fig.~{#1}}

\newcommand{\refeq}[1]{Eq.~({#1})}

\newcommand{\iqsnospace}{IQ-modulators}

\hypersetup
{
bookmarksopen=true,
bookmarks=true, 
pdftitle={},
pdfauthor={Julian B\"{o}hm, Andre Brandst\"{o}tter, Philipp Ambichl, Stefan Rotter, Ulrich Kuhl}, 
pdfsubject=, 
pdftoolbar=true, 
pdfmenubar=true, 
pdfhighlight=/O, 
hyperindex=true, 
colorlinks=true, 
pdffitwindow=true, 
breaklinks=true, 
linkcolor= blue, 
citecolor=blue, 
urlcolor=blue 
}

\begin{document}
\date{\today}

\title{Particlelike scattering states in a microwave cavity}
\author{Julian B\"{o}hm}
\affiliation{Universit\'{e} C\^{o}te d'Azur, CNRS, Institut de Physique de Nice (InPhyNi), France, EU}
\author{Andre Brandst\"{o}tter}
\affiliation{Institute for Theoretical Physics, Vienna University of Technology, A-1040, Vienna, Austria, EU}
\author{Philipp Ambichl}
\affiliation{Institute for Theoretical Physics, Vienna University of Technology, A-1040, Vienna, Austria, EU}
\author{Stefan Rotter}
\affiliation{Institute for Theoretical Physics, Vienna University of Technology, A-1040, Vienna, Austria, EU}
\author{Ulrich Kuhl}
\affiliation{Universit\'{e} C\^{o}te d'Azur, CNRS, Institut de Physique de Nice (InPhyNi), France, EU}

\begin{abstract}
We realize scattering states in a lossy and chaotic two-dimensional microwave cavity which follow bundles of classical particle trajectories. To generate such particlelike scattering states we measure the system's transmission matrix and apply an adapted Wigner-Smith time-delay formalism to it. The necessary shaping of the incident wave is achieved in situ using phase and amplitude regulated microwave antennas. Our experimental findings pave the way for establishing spatially confined communication channels that avoid possible intruders or obstacles in wave-based communication systems.
\end{abstract}

\pacs{42.25.Bs,03.65.Nk,05.45.Mt}

\maketitle

\addcontentsline{toc}{section}{Introduction}
{\textit{Introduction.}--}
The propagation of waves in complex media is widely studied in physics \cite{akkermans2007}. To probe the scattering properties of a medium one typically uses well-defined incident waves and measures their spatial profile at the output. However complex the scattering process may be, the output stays deterministically related to the input such that any change in the input parameters can be directly related to changes in the output pattern. This deterministic relation is encapsulated in a system's scattering matrix \cite{POP10} whose massive information content is exploited through wavefront shaping \cite{VEL07, MOS12, rott16}. The basic idea in this emerging field is to manipulate the incident waves in such a way that a certain output is achieved. This concept was pushed forward due to the possibility of sufficient input control. Spatial light modulators in optics \cite{Gehner2006, vanPutten:08, lueder2010, Maurer2011}, IQ-modulators or spatial microwave modulators in the microwave field \cite{HEN04,KAI14b,BOE16} and transducers in acoustics \cite{FIN97} offer the possibility to use this concept in a large variety of physical disciplines.

Early goals were the development of new schemes for the focusing or defocusing of waves and the compression of pulses \cite{VEL07,LER07,Katz2011, McCabe2011, Weiner2011} \textit{behind} a disordered slab. Special wave patterns can also be achieved \textit{within} such a medium like for states that are spatially focused on an embedded target \cite{Judkewitz2013, Vellekoop2008,AMB16}. In multi-mode fibers also so-called "principal modes" were recently generated that are focused in time both at the input and the output facet of the fiber \cite{CAR15, XIONG16}. States with the unique feature of remaining focused both in space and time during the entire propagation through a complex medium are the so-called particlelike scattering states (PSSs) \cite{ROT11}. These waves form highly collimated beams propagating along the bouncing pattern of classical particles. As a result they avoid multipath interference already by construction leading to an extremely broadband and stable transmission behavior. These features make PSSs ideally suited for the transfer of information in a secure, robust and a directed way without losing part of the transmitted signal to the environment.

\begin{figure}
  \includegraphics[width=\columnwidth]{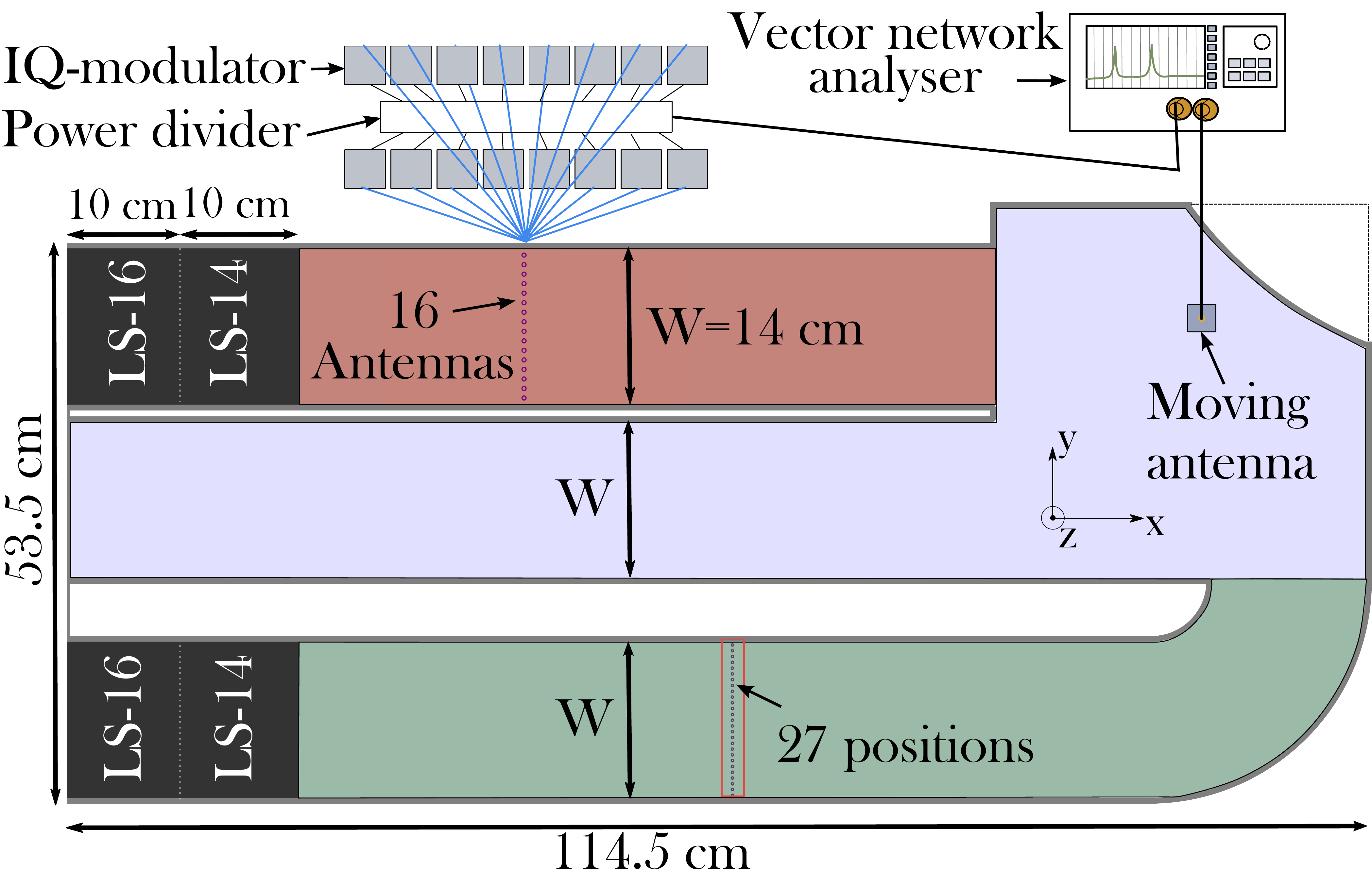}
  \caption{\label{fig:setupwithalllength}
  Sketch of the experimental setup. A two dimensional microwave cavity is excited by 16 monopole antennas (left upper corner) placed in the incoming lead (red colored area). The antennas are connected to IQ-modulators controlling amplitude and phase of the microwave signal, whereas the IQ-modulators themselves are fed by a vector network analyzer. The chaotic scattering region (light blue area) is connected to the incoming lead and an outgoing one (green area). To avoid evanescent coupling to the scattering region, the antennas are placed with a distance of around 2.5 times the lead width $W$ away from the entry to the scattering region. Both leads are closed by absorbing foam material (LS-14, LS-16). The dotted line in the exit lead indicates the 27 positions where the movable monopole antenna is placed for the measurement of the transmission matrix $T$. The red square in the exit lead marks the area for the computation of $I_{\mathrm{ob}}$ and $I_{\mathrm{em}}$ (see text).}
\end{figure}

In this paper we present a microwave realization (see Fig.~\ref{fig:setupwithalllength}) of such PSSs by means of an active input shaping rather than through a numerical synthesis of experimentally measured system excitations as previously reported in \cite{GER16}. Additionally, the realization scheme presented here is solely based on the system's transmission matrix $T$ rather than on the whole scattering matrix $S$ as originally proposed in \cite{ROT11}. Moreover we show here how to deal with the intrinsic losses as well as the noise in our experimental setup.

\addcontentsline{toc}{section}{Theory}
{\textit{Theory.}--}
We start by introducing the Wigner-Smith time-delay matrix (WSTDM) $Q=-iS^{-1}{\text{d}}S/{\text{d}}\omega$ (involving a frequency derivative) which is an established tool for measuring the time-delay associated with the scattering of a wave packet in a system \cite{EIS48,WIG55,SMI60} featuring also interesting connections to the system's density of states \cite{davy2015, Davy2015b, Pierrat2014, savo2017}. The WSTDM is Hermitian for unitary scattering systems, i.e., $S^{\dagger} S= \mathds{1}$, thus resulting in real eigenvalues $\tau_n$ ($n$ represents the $n$-th eigenvalue) also called proper delay-times. The corresponding eigenvectors $\vec{u}_n$ (given as a coefficient vector in a certain basis) are known as principal modes \cite{FAN05} and have the remarkable feature of being insensitive (to first order) to small changes of their input frequency -- in the sense that the spatial output profile $\vec{v}_n=S\vec{u}_n$ does not change (up to a global factor). This property is especially useful for dispersion-free propagation through multi-mode fibers \cite{CAR15, XIONG16, shemirani2009, car2017} and is mathematically expressed as
\begin{equation}\label{eq:static-output}
\vec{v}_n\!\left(\omega_0\!+\! \Delta\omega\right) \!\approx\!
\exp\left({i \tau_n \Delta\omega}\right) \, \vec{v}_n\!\left(\omega_0\right)\!,
\end{equation}
where $\Delta \omega$ is the change in frequency and $\omega_0$ the frequency at which $\vec{u}_n$ is evaluated. The global phase factor $\exp\left({i \tau_n \Delta\omega}\right)$ is determined by the corresponding eigenvalue $\tau_n$. In Ref.~\cite{ROT11} it was demonstrated that a certain subclass of principal modes have a particlelike wave function resembling a focused beam. These PSSs live in the subspace of either fully transmitted or fully reflected states, just like a particle that can either traverse the scattering region or be reflected back at some boundary or obstacle. In this work we investigate PSSs that get fully transmitted through a microwave cavity as shown in Fig.~\ref{fig:setupwithalllength}.

As in most experiments, we also don't have access to the full scattering matrix $S$. We thus use a modified WSTDM where we replace the scattering matrix $S$ by the transmission matrix $T$, which is accessible in our experimental setup shown later. In the following we show that only the knowledge of the transmission matrix $T$ is sufficient to find PSSs connecting the input to the output. The eigenvalue equation for the $n$-th eigenvector $\vec{q}_{n}$ of this new operator $q$ reads as follows:
\begin{equation}
\label{eq:eigvaleqforq}
	 q\,\vec{q}_{n}=-\text{i}\,T^{-1}(\omega)\frac{\text{d}T(\omega)}{\text{d}\omega}\vec{q}_{n}=\lambda_{n}\vec{q}_{n},
\end{equation}
where $\lambda_n$ is the eigenvalue. Please note that an ordinary inverse of $T$ appearing in Eq.~\eqref{eq:eigvaleqforq} does not exist if $T$ is non-quadratic or singular. In the supplemental material \cite{supp}, we introduce an effective inverse that still allows for the calculation of $q$. Contrary to eigenstates of the Hermitian operator $Q$, only the transmitted output profile $\vec{o}_{n}=T\vec{q}_{n}$ is insensitive (up to a global factor) with respect to a change of the input frequency $\omega$, since the operator $q$ involves only the transmission matrix $T$. This translates into
\begin{equation} \label{eq:static-trans-output}
\vec{o}_n\!\left(\omega_0\!+\! \Delta\omega\right) \!\approx\!
\exp\left({i \lambda_n \Delta\omega}\right) \, \vec{o}_n\!\left(\omega_0\right)\!.
\end{equation}
One can analytically derive (see Ref. \cite{FAN05}) an expression for the complex eigenvalues
\begin{eqnarray}
\label{eq:eigvalefullexp}
	\lambda_{n} = \frac{\text{d}\phi_n}{\text{d}\omega}-\text{i}\frac{\text{d}\text{ln}(|{\vec{o}_n}|)}{\text{d}\omega},
\end{eqnarray}
where $\phi_n$ is the transmitted global phase, i.e., $\vec{o}_{n}=|\vec{o}_{n}|e^{i\phi_n}\hat{o}_{n}$ ($\hat{o}_{n}$ is the unit vector of $\vec{o}_{n}$). The real part $\text{Re}(\lambda_{n})$ reflects the frequency derivative of the scattering phase and is therefore proportional to the time-delay \cite{EIS48} of the eigenstate $\vec{q}_{n}$. The imaginary part $\text{Im}(\lambda_{n})$ describes how the transmitted intensity $|{\vec{o}_n}|^2$ changes with respect to a change of the frequency $\omega$ as can be seen from Eq.~\eqref{eq:static-trans-output}, $|\vec{o}_n\!\left(\omega_0\!+\! \Delta\omega\right)|^2 \!\approx\! \exp\left[{-2\text{Im}(\lambda_n) \Delta\omega}\right] \, |\vec{o}_n\!\left(\omega_0\right)\!|^2$. In order to identify PSSs among all the other eigenstates $\vec{o}_{n}$, we make use of the corresponding eigenvalues $\lambda_{n}$. Since PSSs are highly collimated and are not distributed all over the scattering region, the time it takes a PSS to traverse the scattering region is typically much smaller than for other scattering states that get scattered multiple times inside the scattering region. Due to the fact that $\text{Re}(\lambda_{n})$ measures this scattering time, i.e., the time-delay, PSSs can be identified by a small value of $\text{Re}(\lambda_{n})$. Furthermore, PSSs feature a small $\text{Im}(\lambda_{n})$, since for fully transmitting and spatially confined scattering states the transmitted intensity barely changes with input frequency $\omega$ as compared to states that get scattered multiple times. In conclusion, PSSs can be identified by a small $\text{Re}(\lambda_{n})$ and a small $\text{Im}(\lambda_{n})$.

\addcontentsline{toc}{section}{Setup}
{\textit{Setup.}--}
The scattering setup with which we investigate the appearance of PSSs is shown in \reffig{\ref{fig:setupwithalllength}}. A chaotic scattering region is attached to an incoming lead and an outgoing lead (see red, light blue and green areas in \reffig{\ref{fig:setupwithalllength}}, respectively). The width of the lead $W$ of $14\,\text{cm}$ allows the propagation of 16 transverse-electrical modes (TE-modes) in the entrance and the exit lead at the working frequency of $\nu_{0}=\omega_0/2\pi=17.5\,\text{GHz}$ which corresponds to a wavelength in air of $1.71\,$cm. These 16 modes are excitable via 16 antennas. Each antenna is connected to one IQ-modulator, which controls amplitude and phase of the microwave signal passing through. The IQ-modulators themselves are fed by a vector network analyzer (VNA, Agilent E5071C) connected to a power splitter (Microot MPD16-060180). The used connecting cables, connectors and antennas are all identical to avoid the appearance of additional phases occurring from different propagation path lengths. The ends of the waveguide are filled with absorbing foam material (types: LS-14 and LS-16 from EMERSON\,\&\,CUMING) to reduce reflections from the open ends.

The cavity is placed under a metallic plate featuring a $5 \,{\times}\, 5\,\text{mm}^{2}$ grid of holes (hole radius of $2\,$mm). Working below the cut-off frequency of $18.75\,\text{GHz}$ guarantees to excite only the fundamental TE mode, i.e., TE$_{0}$, where the $z$-component of the electric field $E_{z}$ is constant with respect to $z$ and the $x$,$y$-components $E_{x,y}$ are zero. The grid of holes in the top plate closing the cavity enables us to introduce a movable monopole antenna which measures $E_{z}$ at any given hole position in the cavity as in previous experiments \cite{UNT11,BAR13}. The holes can also be used to insert cylindrical obstacles (aluminum, radius: $2\,$mm) leading to additional scattering within the cavity. Similar setups with ten open modes have been used to verify the shaping performance of our antenna array \cite{BOE16} and achieve focusing inside disordered media with a generalized Wigner-Smith matrix \cite{AMB16}.
The scattering setup was chosen such that the incoming and outgoing leads support a sufficient number of modes. The long middle part (located between incoming and outgoing lead) serves as verification that particlelike scattering states avoid this region whereas arbitrary scattering states typically enter this part.

\begin{figure}
  \includegraphics[width=.99\columnwidth]{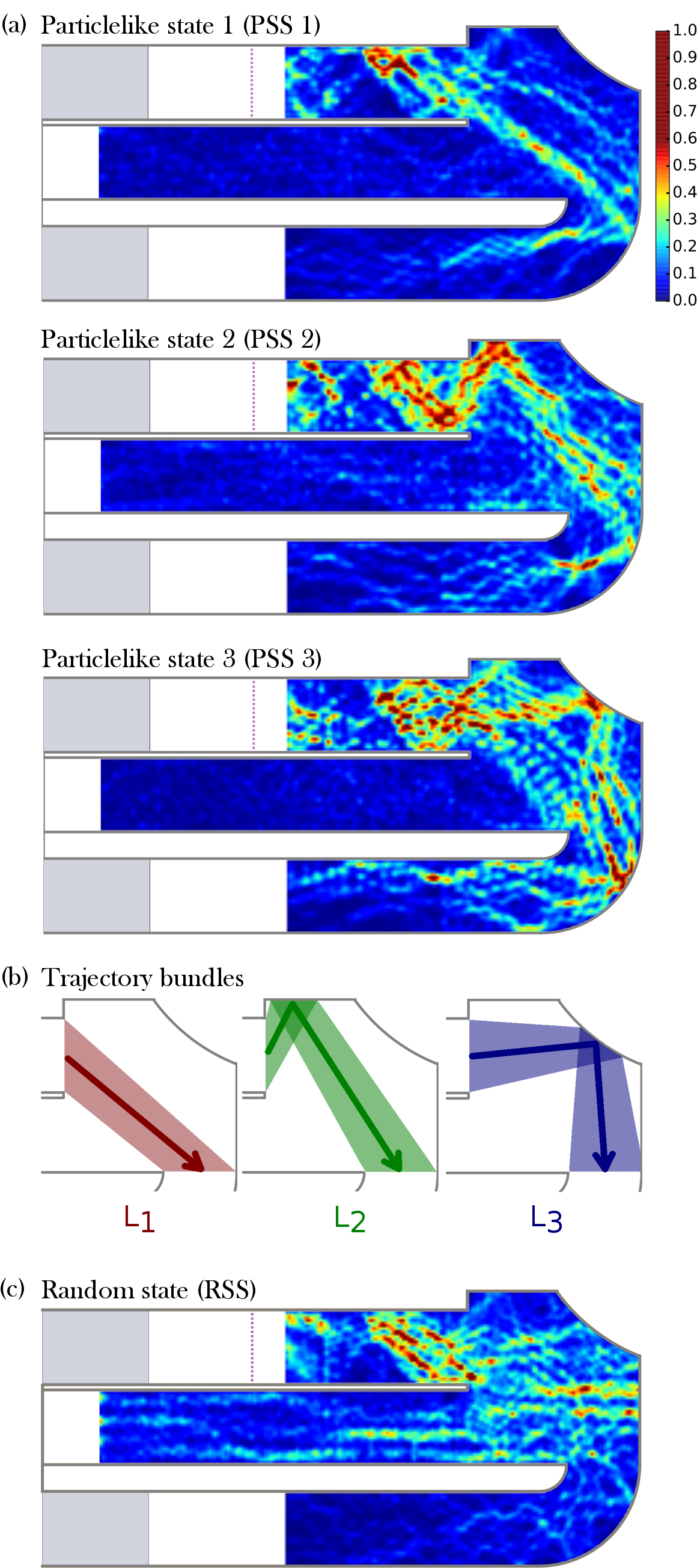}
\caption{\label{fig:intensityparticleLikestatespluspath}
(a) Intensity of the particlelike scattering states which are experimentally created by wavefront shaping of the incident wave in the left upper lead. (b) Central path length of the corresponding classical trajectory bundles of the particlelike scattering states ($L_{1}=33.3\,\text{cm}$, $L_{2}=50.0\,\text{cm}$, $L_{3}=49.1\,\text{cm}$). (c) Typical intensity distribution when only one single antenna is excited (here: 2\textsuperscript{nd} antenna from top).
Note that the intensities in (a) and (c) are normalized each to a maximum (minimum) value of 1 (0).}
\end{figure}

\addcontentsline{toc}{section}{Experimental results}
{\textit{Experimental results.}--}
To obtain the $q$-operator experimentally, we measure the transmission matrix $T(\omega)$ within a frequency window around the working frequency. The positions where we inject the microwave signal (16 antennas connected to the \iqsnospace) and the positions where we measure the transmission with the moving antenna (27 positions) are marked in \reffig{\ref{fig:setupwithalllength}}. Since $T$ is thus a rectangular matrix of size $27\times 16$, we cannot calculate its ordinary inverse to construct the $q$-operator. Using the technique described in the supplemental material \cite{supp} we work with the operator $\tilde{q}$ which only includes a subpart of the full transmission matrix associated to a certain number $\eta$ of highly transmitting channels. We tested empirically that the best results for PSSs [identified by means of a small $\text{Im}(\lambda_n)$] are obtained if we take the highest $\eta=7$ transmitting channels for the calculation of $\tilde{q}$. Once these eigenstates of $\tilde{q}$ are evaluated, we inject them and verify their particlelike shape using the movable antenna that enters the cavity through the holes in the top plate. At first we investigate the eigenstate featuring the smallest value of $\text{Re}(\lambda_n)$, i.e., the shortest time-delay. The result of this measurement is shown as particlelike scattering state 1 (PSS 1) in \reffig{\ref{fig:intensityparticleLikestatespluspath}}(a). The wave function clearly shows the predicted behavior of following the shortest trajectory bundle connecting the incoming with the outgoing lead [see red bundle in \reffig{\ref{fig:intensityparticleLikestatespluspath}(b,left)}].

In the next step we investigate PSSs with larger time-delays, which correspond to the green and the blue classical trajectory bundles shown in \reffig{\ref{fig:intensityparticleLikestatespluspath}(b)}. It turns out that the center trajectories of these two bundles have almost the same length ($L_2=50.0$~cm and $L_3=49.1$~cm). Since similar path lengths lead to similar time-delays, the operator $q$ cannot fully discriminate between these two scattering states. While PSS 2 corresponds quite well to the green classical bundle, PSS 3 mixes both bundles, green and blue. In other words, the measured $q$-eigenstates corresponding to these bundles are in a near-degenerate super-position with path contributions of both lengths ($L_2$ and $L_3$) showing up in their wave functions. Demixing degenerate PSSs can be achieved by transforming the scattering matrix into the spatial domain \cite{GER16}. In order to emphasize the particlelike shape of the PSSs, we also show in \reffig{\ref{fig:intensityparticleLikestatespluspath}(c)} the intensity distribution of a state governed by exciting only a single randomly chosen antenna [we will refer to this state as random scattering state (RSS)] and compare its spatial shape with our PSSs. We see that also the RSS shows some intensity maxima within the cavity, however, these maxima do not follow classical trajectory bundles between incoming and outgoing lead. Moreover, the RSS extends into the middle part of the scattering region which is clearly avoided by all PSSs.

\begin{figure}
  \includegraphics[width=6cm]{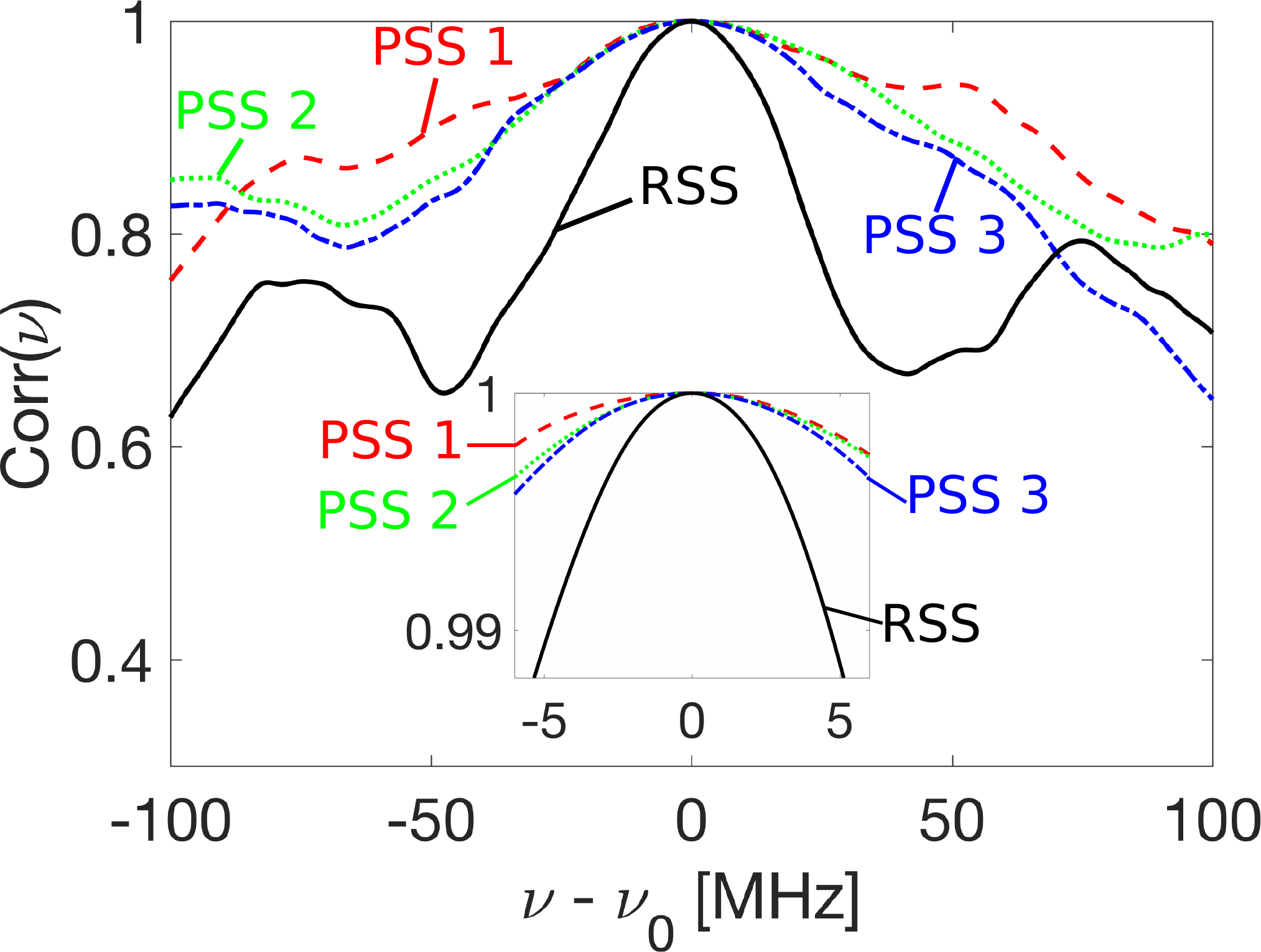}
  \caption{ \label{fig:outputcorrel}
  \noindent {Auto-correlation function of the output profile of the three particlelike scattering states (PSS) and the random scattering state (RSS) according to \refeq{\ref{eq:correl}}. All three PSSs are more stable with respect to a change of the incident frequency $\nu$ as compared to the RSS.}}
\end{figure}

The RSS also shows a significantly lower spectral robustness of its transverse output profile which we define as
\begin{equation}
\label{eq:correl}
\mathrm{Corr}(\nu)=\frac{|\vec{o}^{\,\dagger}(\nu)\cdot \vec{o}(\nu_{0})|}{|\vec{o}(\nu)||\vec{o}(\nu_{0})|} \quad \text{with} \quad \vec{o}(\nu)=T(\nu)\vec{i},
\end{equation}
where $\vec{i}$ is the random input state. Equation \eqref{eq:correl} is the normalized correlation between the output vector $\vec{o}$ at frequency $\nu$ compared to its output at $\nu_{0}$ (the frequency at which the states are evaluated). PSS 1 is the state showing the highest output robustness when compared to the other PSSs (see \reffig{\ref{fig:outputcorrel}}). Since PSS 2 and PSS 3 perform a reflection at the convex cavity boundary, they are considerably more sensitive with respect to small changes in the frequency in terms of the output robustness when compared to PSS 1 which is transmitted entirely without any boundary reflections. This explains why the correlation curve in \reffig{\ref{fig:outputcorrel}} of PSS 1 is flatter than the one of PSS 2 and 3. The correlation of the random state RSS is, as expected, the lowest.

\begin{figure}
  \includegraphics[width=\linewidth]{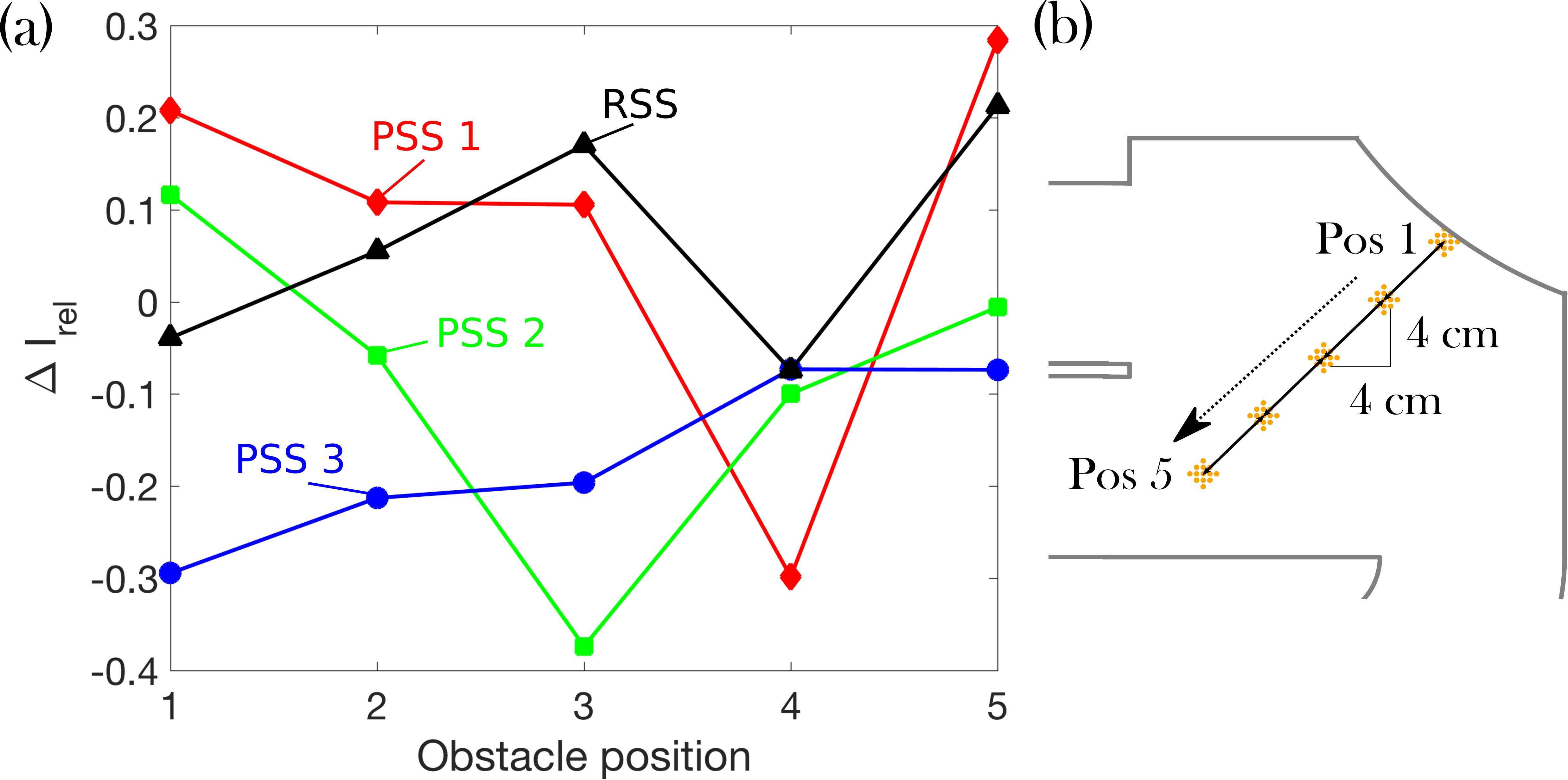}
  \caption{ \label{fig:obstaclesplusmap}
  \noindent {(a) Change of the transmitted intensity $\Delta I_{rel}$ of the particlelike scattering states (PSS) and the random state while moving the rhombic obstacle between position 1 (Pos 1) to position 5 (Pos 5)} [see (b)] operating at the center frequency $\nu_{0}=17.5~$GHz. PSS 1 and PSS 2 show a significant drop at only one position where the obstacle crosses the classical trajectory bundle. As expected, PSS 3, which is a superposition of two classical bundles, is affected by all obstacle positions. The RSS shows no indication of a particlelike pattern.}
\end{figure}

The most characteristic property of PSSs is their highly collimated wave functions occupying bundles of classical particle trajectories of similar length. Consequently, putting an obstacle in the way of such a trajectory bundle, the observed transmission for a corresponding PSS drops down, whereas putting an obstacle outside of the occupied region of the PSS affects the wave function only slightly. To test this idea explicitly experimentally, we place altogether 13 cylindrical obstacles forming a rhombic shape into the scattering region of the cavity [see \reffig{\ref{fig:obstaclesplusmap}}(a)]. In total we place this obstacle at 5 different positions indicated in \reffig{\ref{fig:obstaclesplusmap}(a)} and study the relative change of transmitted intensity according to
\begin{equation}
\label{eq:relativintlossgain}
\Delta I_{\mathrm{rel}}= \frac{(I_{\mathrm{ob}}-I_{\mathrm{em}})}{I_{\mathrm{em}}},
\end{equation}
where $I_{\mathrm{ob}}$ is the transmitted intensity for the case where the obstacle is placed inside the system and $I_{\mathrm{em}}$ is the transmitted intensity for the empty cavity with no obstacle present.
The intensities $I_{\mathrm{ob}}$ and $I_{\mathrm{em}}$ are obtained by computing the sum of the measured transmitted intensities at $135$ positions covering the whole width of the outgoing lead indicated by a red square in the exit lead in \reffig{\ref{fig:setupwithalllength}}.
Following our considerations from above we would expect that the transmitted intensity of a PSS is not affected by an obstacle unless it is placed directly into its corresponding classical trajectory bundle.
Indeed, the PSSs are affected by a strong drop of 30\% or more of the transmitted intensity when the scatterer is placed within the bundle supporting the PSS (see \reffig{\ref{fig:obstaclesplusmap}}). This observation is interesting from a practical point of view if one aims to transmit intensity from input to output lead in the presence of obstacles. Once a specific PSS is blocked, one can maintain efficient transmission by switching to another PSS, e.g., from PSS 1 to PSS 2 or to PSS 3.

\addcontentsline{toc}{section}{Conclusion}
{\textit{Conclusion.}--}
We perform an in situ realization of particlelike scattering states by means of incident wavefront shaping. Particlelike scattering states follow bundles of classical trajectories of similar length and can be identified with the help of the Wigner-Smith time-delay formalism based on a prior measurement of the frequency dependent transmission matrix $T(\nu)$. We extract three different particlelike scattering states corresponding to three different classical trajectory bundles connecting the input with the output lead attached to a chaotic microwave cavity. Switching between these paths can augment the transmission in case one of the paths is blocked by an obstacle. Our results can also be mapped onto other wave based systems (acoustic, electromagnetic, quantum, etc.) leading to many possible applications related to efficient, robust, and focused transmission through complex environments \cite{SAL85,SAM02}. While the presented microwave experiment using 16 guided modes serves as a proof-of-principle demonstration, we expect that our protocol unfolds its full potential in the optical domain where many more modes are accessible.

\addcontentsline{toc}{section}{Acknowledgments}
{\textit{Acknowledgments.}--}
P.A., A.B., and S.R. are supported by the Austrian Science Fund (FWF) through project numbers SFB-NextLite F49-P10 and I 1142- N27 (GePartWave). J.B. and U.K. would like to thank the ANR for funding via the ANR Project GePartWave (ANR-12-IS04-0004-01) and the European Commission through the H2020 programme by the Open Future Emerging Technology "NEMF21" Project (664828).

\section*{Supplementary material}
\subsection{$q$ for non-quadratic or singular transmission matrices $T$}
The construction of $q=-\text{i}\,T^{-1} \text{d}T(\omega)/\text{d}\omega$ involves the inverse of the transmission matrix $T^{-1}$. If $T$ is not quadratic or singular, which can be the case in systems with low transmission, an ordinary inversion cannot be computed anymore. However, an effective inverse can be calculated by using only highly transmitting channels of $T$, as we explain in the following. We start with a singular value decomposition (SVD) of the transmission matrix $T = U \Sigma V^\dagger$, where $U$ consists of the eigenvectors of $TT^{\dagger}$ stored in its columns and $V$ consists of the eigenvectors of $T^{\dagger}T$, respectively. For a $m \times n$-dimensional transmission matrix, the matrices $U$ and $V$ are quadratic $m \times m$ and $n \times n$ matrices. The rectangular $m \times n$-dimensional matrix $\Sigma$ contains the singular values $\sigma_i$ on its diagonal, which are the square roots of the common eigenvalues of both $T^{\dagger}T$ and $TT^{\dagger}$. For a singular or non-quadratic transmission matrix $T$, at least one singular value is zero. In a next step, we only keep a certain number $\eta$ of large singular values $\tilde{\Sigma}$ with corresponding singular vectors stored in $\tilde{U}$ and $\tilde{V}$. Projecting the full transmission matrix $T$ onto the kept transmitting channels according to $\tilde{T}=\tilde{U}^{\dagger}T\tilde{V}$, we end up with an $\eta \times \eta$-dimensional invertible transmission matrix $\tilde{T}$. Projecting back onto the original vector space gives the effective inversion
\begin{equation}
T^{-1}:=\tilde{V}(\tilde{U}^{\dagger}T\tilde{V})^{-1}\tilde{U}^{\dagger}.
\end{equation}
Projecting also the derivative onto the selected subspace with the corresponding projection operators $P_{\tilde{U}}=\tilde{U}\tilde{U}^{\dagger}$ and $P_{\tilde{V}}=\tilde{V}\tilde{V}^{\dagger}$, we end up with the construction rule for the operator
\begin{equation}
\tilde{q}=-i\tilde{V}(\tilde{U}^{\dagger}T\tilde{V})^{-1}\tilde{U}^{\dagger}\tilde{U}\tilde{U}^{\dagger}\frac{\text{d}T}{\text{d}\omega}\tilde{V}\tilde{V}^{\dagger},
\end{equation}
whose eigenvectors can now be calculated readily.

\end{document}